Analytic solutions of the modified Langevin equation in a mean-field model

Yu.A. Koksharov

*Faculty of Physics M.V.Lomonosov Moscow State University, 119991, Moscow, Russia*

*Kotelnikov Institute of Radioengineering and Electronics RAN, 125009, Moscow, Russia.*

**Abstract**

Approximate analytical solutions of the modified Langevin equation are obtained. These solutions are relatively simple and enough accurate. They are illustrated by considering a mean-field model of a system with interacting superparamagnetic particles. Within the framework of this model we derived analytical approximate formulas for the temperature dependencies of the saturation and remnant magnetization, coercive force, initial magnetic susceptibility as well as for the law of approach to saturation. We obtained also some exact analytical relationships for the coercive force. We found remarkable similarity between the approximate cubic equation, which is resulted from the modified Langevin equation, and the exact equation resulting from the divergence condition of a solution derivative. The analytical formulas obtained in this work can be used in various models (not only magnetic ones), where the modified Langevin equation is applied.

**1. Introduction**

The Langevin and Brillouin equations, which are commonly utilized in various mean-field models [1-12], have exact solutions in the form of definite integrals [13, 14]. These integrals are too cumbersome for applications. Hence there is a need in simple analytic solutions, even though these are approximate [3]. The purpose of this work is to obtain such solutions of the modified Langevin equation [2]:

$$z = L(\alpha + \beta z) \quad (1)$$

Here $L(\xi)$ is the Langevin function:

$$L(\xi) = \coth(\xi) - 1/\xi \quad (2)$$

The parameters $\alpha$ and $\beta$ in equation (1) are discussed for a mean-field model in section 4. In this model $\alpha \sim H_0/T$, $\beta \sim \lambda/T$ where $H_0$ is an external magnetic field, $T$ is the absolute temperature, $\lambda$ is the averaged measure of internal magnetic interactions.

Two approximants of (1) are introduced in the next section. Their analytical solutions are discussed in section 3. As will be shown in sections 4-5, these solutions allow to get analytic formulas for temperature dependencies of magnetic parameters in a mean-field model.

## 2. Approximations of modified Langevin equation

There exist quite a few approximate expressions for the inverse Langevin function [15-17]. Formula (3) presents a reasonable compromise between simplicity and accuracy [16, 17]:

$$L^{-1}(z) \approx z\frac{3-z^2}{1-z^2} \qquad (3)$$

The relative error ε of equation (3) can be estimated as

$$\varepsilon = (z - L(z\frac{3-z^2}{1-z^2}))/z \qquad (4)$$

Function (4) reaches its maximal value (≈2%) at z≈±0.65 and vanishes at z = −1; 0; 1. Formula (4) is plotted for z>0 in Fig. 1. Combining (1) with (3) we obtain

$$z\frac{3-z^2}{1-z^2} \approx \alpha + \beta z \qquad (5)$$

As is shown in the next section, roots of equation (5) are quite close to the exact solutions of (1).

The second approximation follows from (2):

$$L(\xi) \approx \pm 1 - \frac{1}{\xi} \qquad (6)$$

for |L(ξ)|≈1, |ξ|>>1.

The relative error ε of equation (6) is given by formula (7):

$$\varepsilon = (L(\xi) - (\pm 1 - 1/\xi))/L(\xi) \qquad (7)$$

The plus and minus sign in (6), (7) corresponds to ξ>0 and ξ<0, respectively.

Formula (7) is plotted in Fig. 1 in the form ε vs z in assuming that ξ>0 and z∈[0,1[, where z≡L(ξ). The relative error of equation (6) equals ≈2% if $L(\xi) \approx 0.6$ and vanishes if $L(\xi) \rightarrow 1$. By the assumption β≠0 we combine (6) with (1) to get equation (8) for α+βz>0 and equation (9) for α+βz<0:

$$z \approx 1 - \frac{1}{\alpha+\beta z} \Rightarrow z^2 + (\frac{\alpha}{\beta}-1)z + \frac{1-\alpha}{\beta} = 0 \Rightarrow$$

$$z = \frac{1}{2}(1-\frac{\alpha}{\beta}) \pm \sqrt{\frac{1}{4}(1+\frac{\alpha}{\beta})^2 - \frac{1}{\beta}} \qquad (8)$$

$$z \approx -1 - \frac{1}{\alpha+\beta z} \Rightarrow z^2 + (\frac{\alpha}{\beta}+1)z + \frac{1+\alpha}{\beta} = 0 \Rightarrow$$

$$z = -\frac{1}{2}(1+\frac{\alpha}{\beta}) \pm \sqrt{\frac{1}{4}(1-\frac{\alpha}{\beta})^2 - \frac{1}{\beta}} \qquad (9)$$

To satisfy |z|<1 one should take the plus sign in (8) and the minus sign in (9).

Formulas (8), (9) are close to the exact solutions of (1) only if |z|>0.6 and, therefore, these approximations are applicable not for all values of α and β, unlike (5). But if |z|≈1, formulas (8) and (9) become much more precise than the solutions of (5), as is shown in the next section.

## 3. The roots of approximate equation (5)

The solutuions of equation (5) are strongly depended on the value of the parameter β. For β≠1, equation (5) can be written as

$$z^3 + Bz^2 + Cz + D = 0, \qquad (10)$$

where

$$B \equiv \frac{\alpha}{\beta-1}; \quad C \equiv -\frac{\beta-3}{\beta-1}; \quad D \equiv -B \qquad (11)$$

Formulas (12)-(15) present the solutuions of cubic equation (10).

$$z = \sqrt{-\frac{4p}{3}} \sin(\frac{\varphi}{3}) - \frac{B}{3} \qquad (12)$$

$$z = \sqrt{-\frac{4p}{3}} \sin(\frac{\varphi+2\pi}{3}) - \frac{B}{3} \qquad (13)$$

$$z = \sqrt{-\frac{4p}{3}} \sin(\frac{\varphi+4\pi}{3}) - \frac{B}{3} \qquad (14)$$

$$z = \sqrt[3]{-\frac{q}{2}+\sqrt{\Delta}} + \sqrt[3]{-\frac{q}{2}-\sqrt{\Delta}} - \frac{B}{3} \qquad (15)$$

Formulas (12)-(17) allow to calculate the real roots z of equation (5) satisfying |z|<1. The formal derivation of (12)-(17) is given in the Appendix. The coefficients p, q, Δ, φ in (12)-(17) are expressed in terms of α and β by equations (A11), (A12), (A14), (A30), respectively (see the Appendix). Each of formulas (12)-(17) is applicable to separate domains in the plane (α, β) (see Fig. 2 and Table 1). There is the smooth joining of functions (12)-(17) on the borders of these domains. The functions $\alpha_{\Delta 1}(\beta)$ and $\alpha_{\Delta 2}(\beta)$ are defined by equations (A15) and (A16), respectively (see the Appendix).

If β=1 then equation (5) reduces to a quadratic one with solutions (16) and (17):

$$z = -\frac{1}{\alpha} - \sqrt{1+\frac{1}{\alpha^2}} \qquad (16)$$

$$z = -\frac{1}{\alpha} + \sqrt{1+\frac{1}{\alpha^2}} \qquad (17)$$

Table 1 shows also the number of the real roots of (5) meeting the requirement $|z|<1$ for the different domains in the plane $[\alpha, \beta]$.

Table 1. The limits of the applicability of formulas (12)-(17).

| Number of roots | β | α | formula |
|---|---|---|---|
| 1 | $]-\infty; 1[$ | $]-\infty;+\infty[$ | (12) |
| 1 | β=1 | $]-\infty;0[$ | (16) |
|   |   | $]0;+\infty[$ | (17) |
| 1 |   | $]-\infty;-\alpha_{\Delta 2}(\beta)[$ | (14) |
| 1 | $]1; 3]$ | $[-\alpha_{\Delta 2}(\beta);\alpha_{\Delta 2}(\beta)]$ | (15) |
| 1 |   | $]\alpha_{\Delta 2}(\beta);+\infty[$ | (13) |
| 1 |   | $]-\infty;-\alpha_{\Delta 2}(\beta)]$ | (14) |
| 1 |   | $[-\alpha_{\Delta 2}(\beta);-\alpha_{\Delta 1}(\beta)]$ | (15) |
|   |   | $[$ | (14) |
| 3 | $]3;+\infty]$ | $[-\alpha_{\Delta 1}(\beta);\alpha_{\Delta 1}(\beta)]$ | (12) |
|   |   |   | (13) |
| 1 |   | $[\alpha_{\Delta 1}(\beta);\alpha_{\Delta 2}(\beta)]$ | (15) |
| 1 |   | $[\alpha_{\Delta 2}(\beta);+\infty[$ | (13) |

It follows from (11) that $B=D=0$ for $\alpha=0$. In this case equation (10) has the trivial solution $z=0$ and the two non-trivial solutions:

$$z = \pm\sqrt{\frac{\beta-3}{\beta-1}} \qquad (18)$$

It is readily verified that $z$ in (20) is real and satisfies the condition $|z|<1$ only for $\beta \geq 3$.

Fig.2 shows that the solution of (5) is unique for $\beta \leq 3$. If $\beta>3$, then equation (5) has the unique solution for $|\alpha|>\alpha_{1\Delta}(\beta)$ or three different solutions for $|\alpha|<\alpha_{1\Delta}(\beta)$. In the case of $|\alpha|=\alpha_{1\Delta}(\beta)$ the two roots out of three coincide.

To demonstrate a peculiarity of equation (5) for $\beta=3$ one can diffirentiate equation (10) with respect to the parameter $\alpha$:

$$3z^2\left(\frac{\partial z}{\partial \alpha}\right)_\beta + 2Bz\left(\frac{\partial z}{\partial \alpha}\right)_\beta + z^2\left(\frac{\partial B}{\partial \alpha}\right)_\beta + C\left(\frac{\partial z}{\partial \alpha}\right)_\beta + z\left(\frac{\partial C}{\partial \alpha}\right)_\beta + \left(\frac{\partial D}{\partial \alpha}\right)_\beta = 0 \quad (19)$$

Substituting (11) into (19), we obtain:

$$\left(\frac{\partial z}{\partial \alpha}\right)_\beta = \frac{1-z^2}{(\beta-1)(3z^2+2Bz+C)} \quad (20)$$

where β≠1.

We conclude that the derivative $\left(\frac{\partial z}{\partial \alpha}\right)_\beta$ is expressed in terms of the function $z$ itself and can be calculated with the help of (12)-(17).

Putting $z=0$ in (20), we can write:

$$\left(\frac{\partial z}{\partial \alpha}\right)_{\beta,\alpha=0} = \frac{1}{3-\beta} \quad (21)$$

Setting (18) in (20), we obtain:

$$\left(\frac{\partial z}{\partial \alpha}\right)_{\beta,\alpha=0} = \frac{1}{(\beta-1)(\beta-3)} \quad (22)$$

where β≥3.

It follows from (21) and (22) that the derivative $\left(\frac{\partial z}{\partial \alpha}\right)_\beta$ of functions (12)-(15) diverges at the point (α, β)=(0, 3).

Let us remark that it is possible to show directly a feature of exact equation (1) at β=3. Indeed, the Langevin function (2) can be written for small ξ:

$$L(\xi) \approx \xi/3, \quad \xi << 1$$

In assuming that α→0 and z→0 we can obtain equation (21) from (1):

$$z \approx \frac{\alpha+\beta z}{3} \quad \Rightarrow \quad z \approx \frac{\alpha}{3-\beta} \quad \Rightarrow \left(\frac{\partial z}{\partial \alpha}\right)_\beta \approx \frac{1}{3-\beta}$$

When β<1, the unique solution of (5) is given by (12) for all values of α. When β=1, the solution is given by (17) for α<0 and (16) for α>0. If α→0 the right-hand side of (16) and (17) tends to zero (see (A6) and (A7) in the Appendix).

For every β∈]1, 3], the unique solution of (5) is given by (15) for |α|≤$\alpha_{\Delta 2}$(β), (14) for α<−$\alpha_{\Delta 2}$(β), and (13) for α>$\alpha_{\Delta 2}$(β).

Figure 3 shows the curves $z(\alpha)|_{\beta=const}$ for different values of β satisfying the condition β<3. Each of these curves is continuous, though, they are calculated by the different formulas (12)-(17) depending on α and β values. The inset (a) presents the two selected curves $z(\alpha)|_{\beta=const}$ on an enlarged scale. The inset (b) shows the relative error of solutions (12)-(17) with reference to numerical solutions of (1) for the three selected curves $z(\alpha)|_{\beta=const}$. As seen in the inset (b), the relative error of formulas (12)-(17) does not exceed 2% except for the point $(\alpha, \beta)=(0, 3)$ at which the derivative $\left(\frac{\partial z}{\partial \alpha}\right)_\beta$ diverges. But the maximal error is rather small ($\approx 3.5\%$) even near this point.

Figure 4 demonstrates in detail the curves $z(\alpha)|_{\beta=2}$ calculated by (13)-(15) and (8)-(9). The insets (a) and (b) show the error of formulas (8)-(9) and (13)-(15), respectively, with reference to the numerical solution of (1), which was obtained by the iteration method [15]. One can see the very high accuracy of (8)-(9) for $|\alpha|>\alpha_{\Delta 2}(\beta)$, where $|z|>0.75$. The error of (13)-(15) does not exceed 2.5% in the whole range of $\alpha$ (see the inset (b)).

The solutions of (5), which are typical for $\beta>3$, are shown in Figs. 5, 6 for $\beta=9$. Pictured together, these solutions form an "S"-shaped curve. The inset (a) in Fig. 5 demonstrates on an enlarged scale the region near the point $\alpha = -\alpha_{\Delta 1}$ at which $\left(\frac{\partial z}{\partial \alpha}\right)_\beta$ diverges. Near this point, as well as near the symmetric point $\alpha = \alpha_{\Delta 1}$, the relative difference between the analytical solution of (5) and the numerical solution of (1) reaches $\approx 10\%$. This difference decreases rapidly away from $\alpha = \pm\alpha_{\Delta 1}$ (see the insets (b), (c) in Fig. 5).

In case of $|\alpha|>\alpha_{\Delta 1}$ the unique solution is given by (14) for $\alpha<-\alpha_{\Delta 2}$, (13) for $\alpha>\alpha_{\Delta 2}$, and (15) for $\alpha_{\Delta 1}<|\alpha|\leq\alpha_{\Delta 2}$. The error of these solutions does not exceed 0.5% (see the inset (c)).

In case of $|\alpha|<\alpha_{\Delta 1}$, one of the three solutions is described by (12). This solution has a negative derivative $\left(\frac{\partial z}{\partial \alpha}\right)_\beta$ and equals to zero at $\alpha=0$. The other solutions, which are given by (13) and (14), have a positive derivative $\left(\frac{\partial z}{\partial \alpha}\right)_\beta$ and equal to $\pm z_R$ at $\alpha=0$, respectively (Fig.6). The value of $Z_R$ is equal to the magnitude of $z$ from (18).

Figure 6 shows in detail the "S"-shaped curve $z(\alpha)$ for $\beta=9$. The inset (a) shows $z_R(\beta)=\sqrt{\frac{\beta-3}{\beta-1}}$ and $z_C(\beta)$, where $z_C(\beta)$ corresponds to the symmetric rounding points of the "S"-shaped curve at which $\left(\frac{\partial z}{\partial \alpha}\right)_\beta \to \infty$. The abscissa of these points equals $\pm\alpha_C(\beta)$, where $\alpha_C(\beta)=\alpha_{\Delta 1}(\beta)$ (see Fig.2). The asymptotic behavior of $\alpha_C(\beta)$ is shown in the inset (c). It follows from A(15) (see the Appendix) that $\alpha_C(\beta)\to\beta$ if $\beta\to\infty$.

To find explicitly $z_C(\beta)$ one can set the denominator in (20) to zero:

$$3z^2 + 2Bz + C = 0 \qquad (23)$$

The quadratic equation (23) has the solution

$$z = -\frac{B}{3} \pm \sqrt{\frac{B^2}{9} - \frac{C}{3}} = \frac{-\alpha \pm \sqrt{\alpha^2 + 3(\beta-3)(\beta-1)}}{3(\beta-1)} \qquad (24)$$

Taking into account $Z_R(\beta) > 0$ we deduce from (24) that

$$z_C(\beta) = \frac{-\alpha_{\Delta 1}(\beta) + \sqrt{\alpha_{\Delta 1}^2(\beta) + 3(\beta-3)(\beta-1)}}{3(\beta-1)} \qquad (25)$$

Substituting $\alpha_{\Delta 1}(\beta)$ from A(15) into (25) we obtain:

$$z_C(\beta) = \frac{\sqrt{2\beta^2 + 6\beta - 9 - (\sqrt{4\beta-3})^3} + \sqrt{(3-4\beta)(3-2\beta+\sqrt{4\beta-3})}}{3\sqrt{2}(\beta-1)} \qquad (26)$$

All segments of the "S"-shaped curve (Fig. 6) are flattened if $\beta$ increases. Indeed, if $\beta \to \infty$, then $z_C(\beta) \to z_R(\beta)$ (see inset (a)), $\gamma_0 \to \pi$ and $\gamma \to 0$ (see inset (c)). The angles $\gamma$ and $\gamma_0$ are shown in Fig. 6. The angle $\gamma$ is well-defined only if $\beta > 3$. The value $tg(\gamma_0)$ and $tg(\gamma)$ is equal to the right-hand side of equation (21) and (22), respectively.

## 4. Mean-field magnetic model

To illustrate the formulas obtained in the previous sections let us consider an ensemble of identical single domain superparamagnetic particles. The particles are randomly dispersed in a solid non-magnetic matrix. The effect of inter-particle magnetic interactions is represented by an effective field $H_{eff}$ that is proportional to the system magnetization $M$.

Neglecting magnetic anisotropy, the magnetization of the superparamagnetic system follows the Langeven law [21]:

$$M = nm = nm_0 L(\vartheta) = nM_0 V_p L(\vartheta), \qquad (31)$$

where

$$\vartheta \equiv \frac{\mu_0 m_0 H}{kT} \qquad (32)$$

is the ratio of the Zeeman's energy $\mu_0 mH$ to the characteristic thermal energy $kT$, $n = N_P/V$ is the volume concentration of the particles, $m_0 = M_0 V_P$ is the particle magnetic moment, $M_0$ is the saturation magnetization of particle material, $V$ is the total system volume, $N_P$ is the number of the particles in the system, $V_P$ is the particle volume, $m$ is thermodynamically averaged magnetic moment of a particle, $H$ is the magnitude of the total magnetic field, $T$ is the absolute temperature, $\mu_0$ is the magnetic permeability of the free space, $k$ is the Boltzmann constant, $L(\vartheta)$ is the Langevin function (2).

The total magnetic field $H$ is equal for all particles in the mean-field model [6]:

$$H = H_0 + \lambda M, \qquad (33)$$

where $H_0$ is a uniform external magnetic field, $\lambda$ is the mean field constant. We suppose that the parameters $n$, $V_P$, $M_0$ in (31) are fixed.

Combining (33) with (31) and (32) we obtain:

$$M = nm_0 L(\frac{\mu_0 m_0 H}{kT}) = nm_0 L(\frac{\mu_0 m_0}{kT}(H_0 + \lambda M)) \qquad (34)$$

Equation (34) can be rewritten as

$$M = xM_0 L(\frac{\mu_0 M_0 V_p H_0}{kT}(1 + \frac{\lambda x M_0}{H_0})), \qquad (35)$$

where $x=nV_p=N_P V_P/V$ is the fraction of the magnetic phase in the system. Using the notations:

$$z = M/xM_0, \qquad (36)$$

$$\alpha = \frac{\mu_0 M_0 V_p H_0}{kT} \qquad (37)$$

$$\beta \equiv \frac{\mu_0 M_0^2 V_p}{kT} \lambda x \qquad (38)$$

we conclude that equation (35) is identical in form to equation (1).

It is convenient to introduce the dimensionless magnetic field

$$h_0 = \frac{\alpha}{\beta} = \frac{H_0}{M_0 \lambda x}, \qquad (39)$$

and the dimensionless temperature

$$\theta = \frac{1}{\beta} = \frac{kT}{\mu_0 M_0^2 V_p \lambda x}. \qquad (40)$$

For simplicity we suppose that $\lambda$ in (39) and (40) is temperature independent. Using (36)-(40), we may rewrite equation (35) as

$$z = L(\frac{h_0 + z}{\theta}) \qquad (41)$$

Equation (41) is similar to equation (4) in [2], which describes hysteresis in ferromagnetic materials.

It follows from (40) that the condition $\beta=3$ corresponds to

$$\theta_c = \frac{1}{3} = \frac{kT_c}{\mu_0 M_0^2 V_p \lambda x} \qquad (42)$$

The critical temperature $T_C$ is equal to

$$T_c = \frac{\mu_0 M_0^2 V_p \lambda x}{3k} \qquad (43)$$

The normalized magnetization z in (41) demonstrates hysteresis behavior (Fig. 5 and 6) if $T<T_C$ ($\theta<\theta_C$, $\beta>3$) and Curie-Weiss behavior if $T>T_C$ ($\theta>\theta_C$, $\beta<3$). Indeed, it follows from (39)-(43) that

$$\left(\frac{\partial z}{\partial H_0}\right)_{T,H_0=0} = \frac{\mu_0 m_0}{3k_B T_C}\left(\frac{\partial z}{\partial h_0}\right)_{T,h_0=0} = \frac{\mu_0 m_0 \beta}{3k_B T_C}\left(\frac{\partial z}{\partial \alpha}\right)_{\beta,\alpha=0} \qquad (44)$$

Combining (21) with (44) and (36) we obtain

$$\chi_0 = \left(\frac{\partial M}{\partial H_0}\right)_{T,H_0=0} = \frac{\mu_0 M_0^2 V_p x}{3k_B T_C}\frac{\beta}{3-\beta} = \frac{\mu_0 M_0^2 V_p x}{3k_B T_C}\frac{T_C}{T-T_C} = \frac{\mu_0 n m_0^2}{3k_B(T-T_C)} \qquad (45)$$

Equation (46) represents the Curie-Weiss law for the initial magnetic susceptibility $\chi_0$ of the system if $T>T_C$.

In experiments one can determine the saturation magnetization of the system $M_s = nm_0 = xM_0$, the critical temperature $T_C$, the initial susceptibility $\chi_0$ as well as the temperature $T$ and the external magnetic field $H_0$. The magnetic moment $m_0$ can be calculated by the use of equation (45):

$$m_0 = \frac{3k_B(T-T_C)\chi_0}{\mu_0 M_S} \qquad (46)$$

The parameters $h_0$ and $\theta$ can be expressed in terms of $H_0$, $T$, $T_C$, which are measured experimentally, and $m_0$, which is calculated by (46):

$$h_0 = H_0\frac{\mu_0 M_0 V}{3k_B T_C} = \frac{\mu_0 m_0 H_0}{3k_B T_C} \qquad (47)$$

$$\theta = \frac{T}{3T_c} \qquad (48)$$

It follows from (39), (40) and (47), (48) that

$$\alpha = \frac{h_0}{\theta} = \frac{\mu_0 m_0 H_0}{k_B T} \qquad (49)$$

$$\beta = \frac{3T_C}{T} \qquad (50)$$

Substituting (49), (50) in (12)-(17) we can find the system magnetization $M$ with reasonable accuracy for all values of $T$ and $H_0$. If $M > 0.65 M_S$ ($z > 0.65$) equations (8) and (9) can be also used with sufficient accuracy. These equations can be written in terms of $h_0$ and $\theta$:

$$z = \frac{1}{2}(1-h_0) + \sqrt{\frac{1}{4}(1+h_0)^2 - \theta} \qquad (51)$$

if $(z+h_0)>0$;

$$z = -\frac{1}{2}(1+h_0) - \sqrt{\frac{1}{4}(1-h_0)^2 - \theta} \qquad (52)$$

if $(z+h_0)<0$. The relative error of (51) and (52) is less than $\approx 1\%$ if $|z|>6.5$ and vanishes when $|z| \to 1$.

If $|h_0| \gg 1$, the law of approach to saturation follows from (51), (52):

$$z \approx \pm 1 \mp \frac{\theta}{1+|h_0|}, \qquad (53)$$

In (53) upper and lower signs correspond to $h_0>0$ and $h_0<0$, respectively. The relative error of (53) is less than $\approx 1\%$ if $|z|>7.5$ and goes to zero if $|z|\to 1$.

## 5. Magnetic hysteresis

Fig. 7 shows the isothermal curve $z(h_0)$, which demonstrates the magnetic hysteresis in the system described in the previous section. If $T<T_C$ ($\beta>3$, $\theta<1/3$) and $-h_C<h_0<h_C$, the magnetization $M$ is not a single-valued function of $h_0$ (Fig. 7). There are three values of $M$, which correspond to the solutions of (5) and calculated by (12)-(14).

The magnetic susceptibility of the solution (12) (line $A_1A_2$ in Fig.7) is negative. This solution is thermodynamically instable [22].

Solutions (13) and (14) (lines $A_1C_1$ and $A_2C_2$ in Fig.7, respectively) have the positive magnetic susceptibility, which can be written, assuming $h_0=0$ and using (22), in the terms $h_0$ and $\theta$ as

$$(\frac{\partial z}{\partial h_0})_{\theta,h_0=0} = \frac{\theta}{3(\theta_C-\theta)(1-\theta)}, \qquad (54)$$

The initial susceptibility (54) is positive, if $\theta<\theta_C$, tends to zero, if $\theta\to 0$ (to meet the third law of thermodynamics) and diverges, if $\theta\to\theta_C$.

The isothermal magnetic susceptibility, which is related to solutions (13) and (14), diverges at the points $A_1$ and $A_2$, where tangential lines to the curves $A_1C_1$ and $A_2C_2$ become vertical (Fig.7).

Abscissae of $A_1$ и $A_2$ meet the condition $-h_C$ and $+h_C$, respectively, where $h_C=\alpha_{\Delta1}/\beta=\theta\alpha_{\Delta1}$. The divergence of the susceptibility at $A_1$ and $A_2$ can be proved directly by substitution of (A34) and (A37) into (20).

The magnetic hysteresis loop can be described as follows. When the magnetic field $h_0$ changes from a maximal positive value (the point $C_1$ in Fig.7) down to $-h_C$, the instability at point $A_1$, which is due to the divergence of the susceptibility, results in the transition $A_1\to B_2$ to the stable state $B_2$. If $h_0$ still decreases to limiting negative value (the point $C_2$ in Fig.7) the magnetization changes smoothly. Increase in $h_0$ from $C_2$ to $A_2$ is accompanied by the continuous change of the magnetization (Fig.7), which is interrupted with the transition $A_2\to B_1$ because of the susceptibility divergence at $A_2$. The uninterrupted portion of the magnetization curve $B_1C_1$ completes the hysteresis loop.

The loop width equals $2h_C$, where $h_C$ may be called the coercive force. The coercive force vanishes at $\theta=\theta_C$ and approaches to 1 at $\theta=0$. The law for the temperature dependence $h_C$ is presented by equation (55), which is derived from (A15):

$$h_C = \sqrt{\frac{2+6\theta-9\theta^2 - (4\theta^{1/3} - 3\theta^{4/3})^{3/2}}{2}} \qquad (55)$$

The argument of the square root in (55) is non-negative if $\theta \leq \theta_C$. The temperature dependence (55) is plotted in Fig.8.

It is possible to obtain simple approximations for equation (54). The first order Taylor expansion results in

$$h_C \approx \sqrt{6}(\theta_C - \theta)^{3/2} \qquad (56)$$

if $\theta \to \theta_C$, and

$$h_C \approx 1 - \frac{2}{\sqrt{3}}\sqrt{\frac{\theta}{\theta_C}} \qquad (57)$$

if $\theta \to 0$.

Numerical calculations shows that in the interval $\theta/\theta_C \in [\theta_S, \theta_F]$, where $\theta_S=2/5$, $\theta_F=4/5$, equation (55) is approximately given by

$$h_C(\theta) = h_F + (h_S - h_F)\left(\frac{1-\theta/\theta_F}{1-\theta_S/\theta_F}\right)^{5/4}, \qquad (58)$$

where

$$h_S = 1 - \frac{2}{\sqrt{3}}\sqrt{\frac{\theta_S}{\theta_C}} = 1 - \sqrt{\frac{8}{15}} \approx 0.27 \qquad (59)$$

$$h_F = \sqrt{6}(\theta_C - \theta_F)^{3/2} = \sqrt{6}(\theta_C/5)^{3/2} \approx 0.04 \qquad (60)$$

The inset in Fig. 8 shows the relative error of the approximations (58)-(60) with reference to the numerical solution of (1). The limiting percentage error equals $\approx 5.5\%$ at $\theta=\theta_F$.

Equation (56) predicts that $dh_C/d\theta \sim (\theta-\theta_C)^{1/2} \to 0$, if $\theta \to \theta_C$. It is interesting that experiments [25-27] indeed show "$\sqrt{T}$"-dependence (similar to (57)) for $h_C(T)$ at low temperatures and the flattening of $h_C(T)$ at high temperatures.

The temperature dependence of the remnant magnetization $z_R$ may be obtained from (18):

$$z_R = \sqrt{\frac{1-3\theta}{1-\theta}} = \sqrt{\frac{1-\theta/\theta_C}{1-\theta}} \qquad (61)$$

The formula (61) is plotted in inset (a) of Fig. 7. The relative error of (61) with reference to the numerical solution of (1) does not exceed $\approx 5\%$, as can be seen from the inset (b) of Fig. 7. In the vicinity of $\theta_C$ equation (61) shows that $z_R \sim (1-\theta/\theta_C)^{1/2}$. The same temperature dependence is typical of quantum mechanical mean-field models [28].

## 6. The exact value of the coercive force

Let us estimate the exact value of the coercive force. For this purpose a cubic equation similar to (5) can be written by differentiation of exact equation (1) in assuming that $\left(\frac{\partial z}{\partial \alpha}\right)_\beta \to \infty$.

It follows from (1) that

$$\frac{dz}{d\alpha} = \frac{dL}{d\xi}\frac{d\xi}{d\alpha} = \frac{dL}{d\xi}(1+\beta\frac{dz}{d\alpha}), \qquad (62)$$

where $\xi=\alpha+\beta z$. Gathering like terms in (62) and using the identity

$$\frac{dL}{d\xi} = 1 - L(L+2/\xi) \qquad (63)$$

we obtain

$$\frac{dz}{d\alpha} = \frac{\frac{dL}{d\xi}}{1-\beta\frac{dL}{d\xi}} = \frac{1-L(L+\frac{2}{\xi})}{1-\beta+\beta L(L+\frac{2}{\xi})} \qquad (64)$$

The condition $\left(\frac{\partial z}{\partial \alpha}\right)_\beta \to \infty$ suggests that the denominator in (64) equals zero:

$$1-\beta+\beta L(L+\frac{2}{\alpha+\beta z}) = 0 \qquad (65)$$

Setting (1) in (65), we obtain the cubic equation:

$$z^3 + B'z^2 + C'z + D' = 0 \qquad (66)$$

where

$$B' \equiv \frac{\alpha}{\beta}; \quad C' \equiv -\frac{\beta-3}{\beta}; \quad D' \equiv -B'(1-\frac{1}{\beta}) \qquad (67)$$

The coefficients (67) can be expressed in terms of the coefficients (11):

$$B' = \frac{\beta-1}{\beta}B; \quad C' = \frac{\beta-1}{\beta}C; \quad D' = D(\frac{\beta-1}{\beta})^2 \qquad (68)$$

Assuming $\beta>3$ ($\theta<\theta_C$), the solutions of (66) can be calculated using equations (12)-(15) and Table 1 after replacement:

$$B \to B'; C \to C'; D \to D'; \quad \alpha_{\Delta 1} \to \alpha'_{\Delta 1}; \alpha_{\Delta 2} \to \alpha'_{\Delta 2} \qquad (69)$$

where

$$\alpha'_{\Delta 1} = \sqrt{\frac{\beta}{\beta-1}}\alpha_{\Delta 1} \qquad (70)$$

$$\alpha'_{\Delta 2} = \sqrt{\frac{\beta}{\beta-1}}\alpha_{\Delta 2} \qquad (71)$$

The solutions of equation (66) and exact equation (1) are identical only at the points, where $\left(\frac{\partial z}{\partial \alpha}\right)_\beta$ for (1) diverges. One such point is shown in Fig. 9 by the arrow. It follows from (66) and (67) that

$$z_0^3 - \frac{\alpha_{c0}}{\beta} z_0^2 - \frac{\beta-3}{\beta} z_0 + \frac{\alpha_{C0}(\beta-1)}{\beta^2} = 0, \qquad (72)$$

where the parameter $\alpha_{C0}$ and magnetization $z_0$ is corresponds to the exact value of the coercive force $h_{C0}$. Replacing $\frac{\alpha_{CO}}{\beta}$ by $h_{C0}$ and $\beta$ by $1/\theta$ in (72), we obtain

$$z_0^3 - h_{C0} z_0^2 - (1-3\theta) z_0 + h_{C0}(1-\theta) = 0 \qquad (73)$$

This is the accurate relationship between the coercive force $h_{C0}$ and the normalized magnetization $z_0$ at the "rounding" point of the hysteresis loop for equation (1) (Fig. 9).

The value of $h_C$ in Fig. 9 is determined by (55). Combining (55) and (70), we obtain also the formula (74) for $h'_C$:

$$h'_C = \sqrt{\frac{2 + 6\theta - 9\theta^2 - (4\theta^{1/3} - 3\theta^{4/3})^{3/2}}{2(1-\theta)}} \qquad (74)$$

The value of $h_{C0}$ may not be larger than $h'_C$ because the solutions of (66) are confined between $-h'_C$ and $+h'_C$. Assuming $h_{C0} < h_C$ we should admit (see Fig. 9) that the curves representing the exact solutions of (1) and the approximate solutions of (5) may cross each other. Such intersection is possible only for the trivial solutions, which are thermodynamically unstable if $\theta < \theta_C$ and should be ignored. Therefore $h_{C0} > h_C$ and we can write

$$\sqrt{\frac{2 + 6\theta - 9\theta^2 - (4\theta^{1/3} - 3\theta^{4/3})^{3/2}}{2(1-\theta)}} > h_{C0} > \sqrt{\frac{2 + 6\theta - 9\theta^2 - (4\theta^{1/3} - 3\theta^{4/3})^{3/2}}{2}} \qquad (75)$$

If $\theta \to 0$ then $h_{C0}$ is practically equals $h_C$ (see the inset in Fig.8)

## 7. Conclusions

The approximate analytical solutions (8), (9), (12)-(17), which are obtained in this paper for the modified Langevin equation (1), are rather simple and enough accurate. An application of these solutions to the magnetic mean-field model results in formulas (61), (51)-(52) and (55), for the temperature dependencies of the saturation and remnant magnetization as well as the coercive force, respectively. We found also the exact formula (73) for the coercive force. We certainly realize that the mean-field model, described in sections 4-6, is too rough to explain complex properties of real superparamagnetic systems. For example, equation (35) neglects of a particle magnetic anisotropy and polydispersivity of such systems. The nonzero magnetic anisotropy results in drastic slowing relaxation of magnetization below the blocking temperature

$T_B$ [23]. To a certain degree neglect of the anisotropy can be justified only above $T_B$ since in the system of superparamagnetic particles with random axis distribution the average magnetization is hardly subject to the uniaxial magnetic anysotropy (see Fig. 11 in [24]). If $T_B \ll T_C$, where $T_C$ is the transition temperature due to magnetic interactions between particles, the formulas of sections 4-6 can be also applied in the vicinity of $T_C$. Polydispersivity of real nanoparticle systems can be considered by introducing some reasonable distribution of magnetic parameters ($M_0$, $V_p$, $\lambda$, etc.). We suppose that analytical formulas derived in this paper can be useful in various mean-field models of ferromagnets [2, 12], ferrocolloids [29-31] and other theories, in which the modified Langevin equation is included.

**Acknowledgements**

The work was carried out within the framework of the state task and partially supported by the Russian Foundation for Basic Research, the project no.18-29-02080.

**Appendix**

If $\alpha=0$, then equation (5) may be written as

$$z\frac{3-z^2}{1-z^2} = \beta z \tag{A1}$$

Equation (A1) has the trivial solution $z=0$. If $z \neq 0$, then we can rewrite (A1) as

$$\frac{3-z^2}{1-z^2} = \beta \tag{A2}$$

It follows from (A2) that

$$z^2 = \frac{3-\beta}{1-\beta} \tag{A3}$$

This enables us to write the non-trivial solutions of (A1) as

$$z = \pm\sqrt{\frac{\beta-3}{\beta-1}} \tag{A4}$$

It follows from (A4) that $|z|<1$ only if $\beta \geq 3$.

It is possible to obtain an approximate equation like (A4) directly from equation (1) in assuming that $\beta(\partial z/\partial \alpha)_\beta \gg 1$ and $\alpha \to 0$:

$$(\frac{\partial z}{\partial \alpha})_\beta = \frac{dL}{d\xi}(\frac{\partial \xi}{\partial \alpha})_\beta = [1 - L(L+\frac{2}{\xi})][1 + \beta(\frac{\partial z}{\partial \alpha})_\beta] \approx$$

$$\approx [1 - z^2 - \frac{2}{\beta}]\beta(\frac{\partial z}{\partial \alpha})_\beta \Rightarrow 1 = [1 - z^2 - \frac{2}{\beta}]\beta \Rightarrow z = \pm\sqrt{1-\frac{3}{\beta}}$$

where $\xi=\alpha+\beta z$. This equation is less accurate than (A4).

If β=1, then equation (5) may be written as

$$\alpha z^2 + 2z - \alpha = 0 \tag{A5}$$

If α=0, then equation (A5) has the unique trivial solution z=0. This is in agreement with the solutions of (A1).

If α≠0, then equation (A5) has the unique non-trivial solution

$$z = -\frac{1}{\alpha} \pm \sqrt{1 + \frac{1}{\alpha^2}} \tag{A6}$$

Equation (A6) requires "+" for α>0 and "−" for α<0 to satisfy |z|<1.

If α→0, then (A6) may be written as

$$z = -\frac{1}{\alpha} \pm \frac{1 + 0.5\alpha^2}{|\alpha|} = \frac{\alpha}{2} \tag{A7}$$

Equation (A7) can be obtained directly from (1) provided that β=1, α→0, z→0:

$$z = L(\alpha + z) \approx \frac{\alpha + z}{3} \Rightarrow z = \frac{\alpha}{2}$$

If β≠1, then we may write (5) as the cubic equation

$$z^3 + Bz^2 + Cz + D = 0, \tag{A8}$$

where

$$B \equiv \frac{\alpha}{\beta - 1}; \quad C \equiv -\frac{\beta - 3}{\beta - 1}; \quad D \equiv -B \tag{A9}$$

Equation (A8) can be reduced to

$$y^3 + py + q = 0, \tag{A10}$$

where

$$p = C - \frac{B^2}{3} = -\frac{\alpha^2 + 3(\beta^2 - 4\beta + 3)}{3(\beta - 1)^2} = -\frac{\alpha^2 + 3(\beta - 3)(\beta - 1)}{3(\beta - 1)^2} \tag{A11}$$

$$q = \frac{2B^3}{27} - \frac{BC}{3} + D = \frac{2\alpha(\alpha^2 - 9\beta(\beta - 1))}{27(\beta - 1)^3} \tag{A12}$$

$$y = z + \frac{B}{3} \tag{A13}$$

Solutions of (A10) depend on the sign of the discriminant Δ [18, 19]:

$$\Delta = \frac{q^2}{4} + \frac{p^3}{27} = -\frac{\alpha^4 + (\beta - 3)^3(\beta - 1) + \alpha^2(9 - 6\beta - 2\beta^2)}{27(\beta - 1)^4} \tag{A14}$$

The equality Δ=0 holds true if α=±$\alpha_{\Delta 1}$ or α=±$\alpha_{\Delta 2}$, where

$$\alpha_{\Delta 1} = \sqrt{\frac{2\beta^2 + 6\beta - 9 - (4\beta - 3)^{3/2}}{2}} \tag{A15}$$

$$\alpha_{\Delta 2} = \sqrt{\frac{2\beta^2 + 6\beta - 9 + (4\beta - 3)^{3/2}}{2}} \tag{A16}$$

It follows from (A11) that $p=0$ if $\alpha=\pm\alpha_p$, where

$$\alpha_p = \sqrt{3(1-\beta)(\beta-3)} \tag{A17}$$

Setting $q=0$ in (A12), we obtain $\alpha=\pm\alpha_q$, where

$$\alpha_q = 3\sqrt{\beta(\beta-1)} \tag{A18}$$

Figure A shows the curves, which are specified by (A15)-(A18) in the plane $(\alpha, \beta)$.

Equation (A10) is simplified if one or both of $p$ and $q$ equals zero. Let us consider all possible variants.

**(1) p=0; q=0**

The equation has the unique trivial solution $z=0$ corresponding to the points (0, 1) и (0, 3) in plane $(\alpha, \beta)$ (Fig. A).

**(2) p=0; q≠0**

Equation (A10) has the unique non-trivial solution

$$y_1 = y_2 = y_3 = -\sqrt[3]{q} \tag{A19}$$

Combining (A12), (A13), (A17), (A19) we obtain

$$y_1 = y_2 = y_3 = \frac{\alpha}{3(\beta-1)}\sqrt[3]{1+\frac{9(\beta-1)}{\beta-3}} \tag{A20}$$

$$z_1 = z_2 = z_3 = \frac{\alpha}{3(\beta-1)}\{\sqrt[3]{1+\frac{9(\beta-1)}{\beta-3}}-1\}, \tag{A21}$$

where $\alpha=\pm\alpha_p$.

If $\beta\rightarrow 1$, then (A21) coincides with (A7):

$$z_1 = z_2 = z_3 \approx \frac{\alpha}{3(\beta-1)}\{1+\frac{9(\beta-1)}{3(\beta-3)}-1\} = \frac{\alpha}{2}$$

**(3) q=0; p≠0**

Equation (A10) has the three roots

$$y_1 = 0;\ y_2 = \sqrt{-p};\ y_3 = -\sqrt{-p} \tag{A22}$$

If $p>0$, then $y_2$ and $y_3$ are imaginary. Hence $y_1$ is the unique real solution of (A10) at the points $(\alpha, \beta)$ with $\alpha=0$, $1<\beta<3$ (see Fig. A).

If $p<0$, then all roots (A22) are real and distinct. This case corresponds to lines $\pm\alpha_q(\beta)$ and the points with $\alpha=0$, $3\leq\beta<\infty$ in Fig. A. The corresponding solutions of (A8) can be deduced from (A22) by (A13)

$$z_1 = -\frac{B}{3};\ z_2 = \sqrt{-p}-\frac{B}{3};\ z_3 = -\sqrt{-p}-\frac{B}{3} \tag{A23}$$

If $\alpha=0$, then it follows from (A23) that

$$z_1 = y_1 = 0; z_2 = y_2 = \sqrt{\frac{\beta-3}{\beta-1}}; z_3 = y_3 = -\sqrt{\frac{\beta-3}{\beta-1}} \tag{A24}$$

If $3 \leq \beta < \infty$, then all solutions (A24) complies with $|z|<1$.

According to (A9), (A11), (A18), and (A24) for the points on the curves $\pm\alpha_q(\beta)$ (Fig. A)

$$z_1 = -\sqrt{\frac{\beta}{\beta-1}}; z_2 = \frac{\sqrt{4\beta-3}-\sqrt{\beta}}{\sqrt{\beta-1}}; z_3 = \frac{-\sqrt{4\beta-3}-\sqrt{\beta}}{\sqrt{\beta-1}} \tag{A25}$$

The only $z_2$ in (A25) complies with $|z|<1$.

**(4) p≠0; q≠0**

Equation (A10) has one real root and two imagine ones if $\Delta>0$, three real roots, two of which are coincide, if $\Delta=0$, and three distinct real roots if $\Delta<0$ [19]. The plane ($\alpha$, $\beta$) contains three connected domains, two of which have $\Delta<0$, and the third has $\Delta>0$ (Fig. A).

If $\Delta>0$, then the unique real root of (A10) is calculated by the Cardano formula

$$y = \sqrt[3]{-\frac{q}{2}+\sqrt{\Delta}} + \sqrt[3]{-\frac{q}{2}-\sqrt{\Delta}} \tag{A26}$$

The root of (A11), corresponding to (A26), is written as

$$z = \sqrt[3]{-\frac{q}{2}+\sqrt{\Delta}} + \sqrt[3]{-\frac{q}{2}-\sqrt{\Delta}} - \frac{B}{3} \tag{A27}$$

The three real roots of (A10) corresponding to the condition $\Delta=0$, which holds for the curves $\pm\alpha_{\Delta 1}(\beta)$ и $\pm\alpha_{\Delta 2}(\beta)$ in Fig. A1, can be written as

$$y_1 = y_2 = \sqrt[3]{\frac{q}{2}}; y_3 = -2\sqrt[3]{\frac{q}{2}} \tag{A28}$$

The root $y_3$ in (A28) can be derived from (A26) by $\Delta\to 0$, i.e. in moving in the plane ($\alpha$, $\beta$) (Fig. A1) from points inside the domain with $\Delta>0$ to the curves $\pm\alpha_{\Delta 1}(\beta)$ or $\pm\alpha_{\Delta 2}(\beta)$. It will be shown (see (A33)) that the roots (A28) can be found also in going to the curves $\pm\alpha_{\Delta 1}(\beta)$ or $\pm\alpha_{\Delta 2}(\beta)$ from points inside the domains with $\Delta<0$.

If $\Delta<0$, then equation (A10) has the three real distinct roots:

$$y_k = \sqrt{-\frac{4p}{3}}\sin(\frac{\varphi+2k\pi}{3}), \tag{A29}$$

where k=0, 1, 2.

The angle $\varphi$ in (A29) may be calculated by the formula

$$\varphi = \arcsin(\frac{9q}{4p^2}\sqrt{-\frac{4p}{3}}) \tag{A30}$$

Note that $p<0$ in (A29) and (A30). Indeed, it follows from (A14) that if $q\neq 0$ and $\Delta\leq 0$, then $p<0$. If $q=0$, then $\Delta$ and $p$ have the same sign.

The roots (A29) meet the condition $y_1+y_2+y_3 = 0$. In view of (A13) we have for the corresponding roots of (A8)

$z_1+z_2+z_3 = -B$

According to (A14), if $\Delta \to 0$ and $q \neq 0$, $p \neq 0$, then

$$q^2/p^3 \to (-4/27) \tag{A31}$$

In view of (A31) equation (A30) is rewritten as

$$\varphi \to \arcsin(\text{sgn}(q)\frac{9}{4}\sqrt{-\frac{4q^2}{3p^3}}) = \text{sgn}(q)\arcsin(\frac{9}{4}\sqrt{\frac{16}{81}}) = \text{sgn}(q)\frac{\pi}{2} \tag{A32}$$

It is evident from Fig. A1 that q<0 and q>0 hold for the curves $+\alpha_{\Delta 1}(\beta)$, $+\alpha_{\Delta 2}(\beta)$ and $-\alpha_{\Delta 1}(\beta)$, $-\alpha_{\Delta 2}(\beta)$, respectively. Hence, in moving from the domains with $\Delta<0$ to the curves $+\alpha_{\Delta 1}(\beta)$ и $+\alpha_{\Delta 2}(\beta)$ the roots (A29) tend to their limit values as follows:

$$y_1 = y_3 \to -\sqrt{-\frac{p}{3}}; \; y_2 \to 2\sqrt{-\frac{p}{3}} \tag{A33}$$

Combining (A11), (A13), (A15), (A16), (A33) we may write the related roots of (A8) as follows:

$$z_1 = z_3 \to -\frac{\sqrt{8\beta^2 - 18\beta + 9 \pm (4\beta-3)^{3/2}} + \sqrt{2\beta^2 + 6\beta - 9 \pm (4\beta-3)^{3/2}}}{3\sqrt{2}(\beta-1)} \tag{A34}$$

$$z_2 \to \frac{2\sqrt{8\beta^2 - 18\beta + 9 \pm (4\beta-3)^{3/2}} - \sqrt{2\beta^2 + 6\beta - 9 \pm (4\beta-3)^{3/2}}}{3\sqrt{2}(\beta-1)}, \tag{A35}$$

where the plus and minus sign should be taken for $+\alpha_{\Delta 2}(\beta)$ and $+\alpha_{\Delta 1}(\beta)$, respectively.

In moving from the domains with $\Delta<0$ to the curves $-\alpha_{\Delta 1}(\beta)$ и $-\alpha_{\Delta 2}(\beta)$ the roots (A29) behave as follows:

$$y_1 = y_2 \to \sqrt{-\frac{p}{3}}; \; y_3 \to -2\sqrt{-\frac{p}{3}} \tag{A36}$$

Using (A11), (A13), (A15), (A16), (A36) we have the formulas for the related roots of (A8):

$$z_1 = z_2 \to \frac{\sqrt{8\beta^2 - 18\beta + 9 \pm (4\beta-3)^{3/2}} + \sqrt{2\beta^2 + 6\beta - 9 \pm (4\beta-3)^{3/2}}}{3\sqrt{2}(\beta-1)} \tag{A37}$$

$$z_3 \to -\frac{2\sqrt{8\beta^2 - 18\beta + 9 \pm (4\beta-3)^{3/2}} - \sqrt{2\beta^2 + 6\beta - 9 \pm (4\beta-3)^{3/2}}}{3\sqrt{2}(\beta-1)}, \tag{A38}$$

where the plus and minus sign are used for $-\alpha_{\Delta 2}(\beta)$ and $-\alpha_{\Delta 1}(\beta)$, respectively.

A comparison between (A34) and (A37), (A35) and (A38) shows that the roots of (A8) satisfy $z(\alpha)=-z(-\alpha)$ for the curves $\pm\alpha_{\Delta 1}(\beta)$ и $\pm\alpha_{\Delta 2}(\beta)$. Using (A31), one can proof the coincidence of (A33), (A36) with (A28).

Algebraic manipulations, which are direct but cumbersome, demonstrate that every root (A34), (A35), (A37), (A38) satisfies the condition |z|<1 for $\pm\alpha_{\Delta 1}(\beta)$ and the only two roots (A35), (A38) do it for $\pm\alpha_{\Delta 2}(\beta)$.

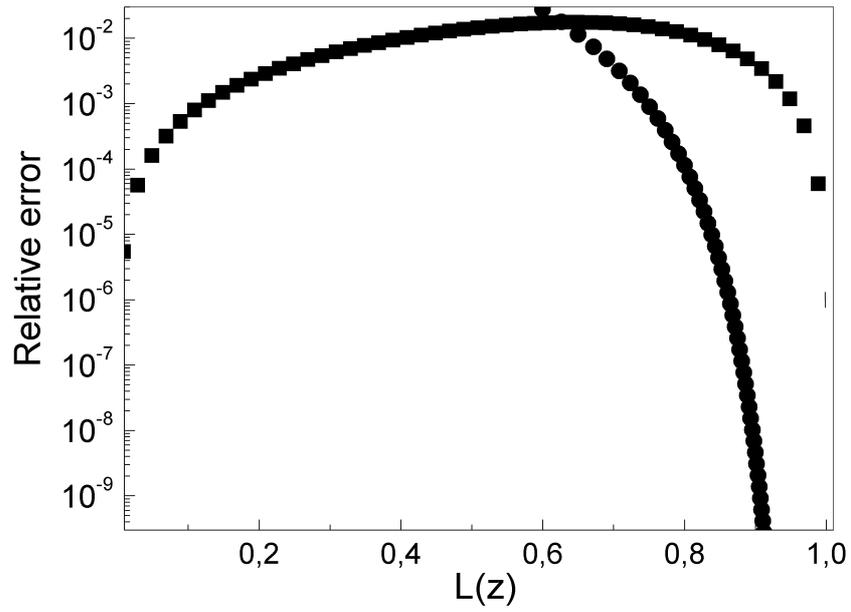

Fig.1 The relative error ε of approximate equations (3) (squares) and (6) (circles) plotted from equations (4) and (7), respectively. In case of (7) we suppose that $\xi>0$ and $z \equiv L(\xi) \in [0,1[$.

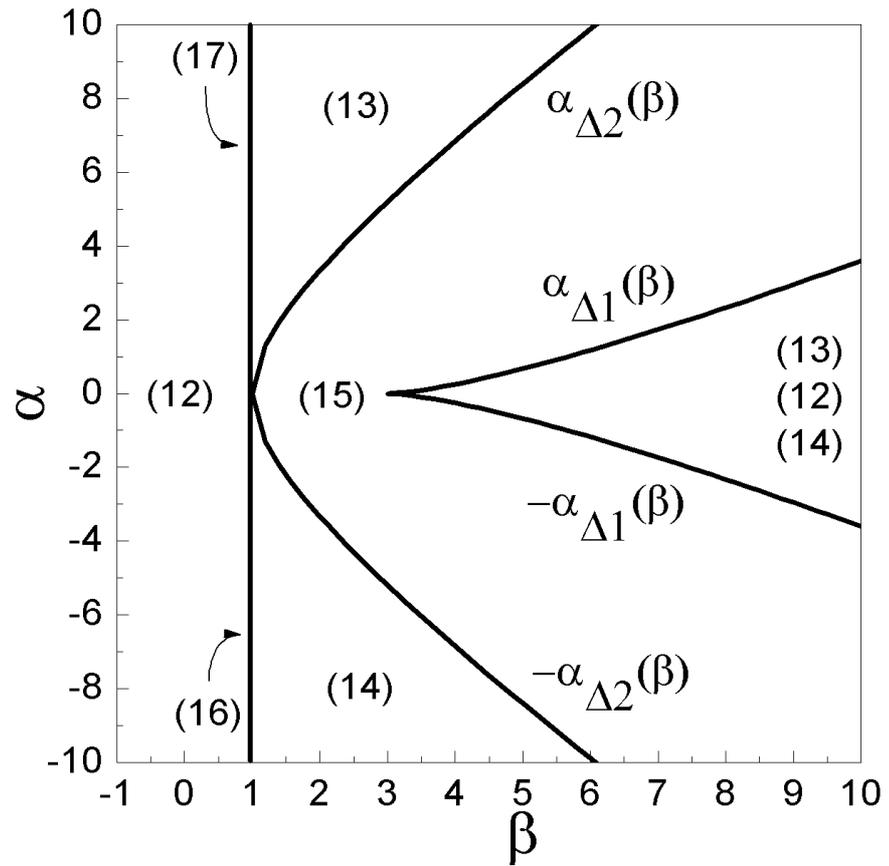

Fig. 2 The applicability regions of equations (12)-(17). The functions $\alpha_{\Delta 1}(\beta)$ and $\alpha_{\Delta 2}(\beta)$ are given by (A15) and (A16), respectively.

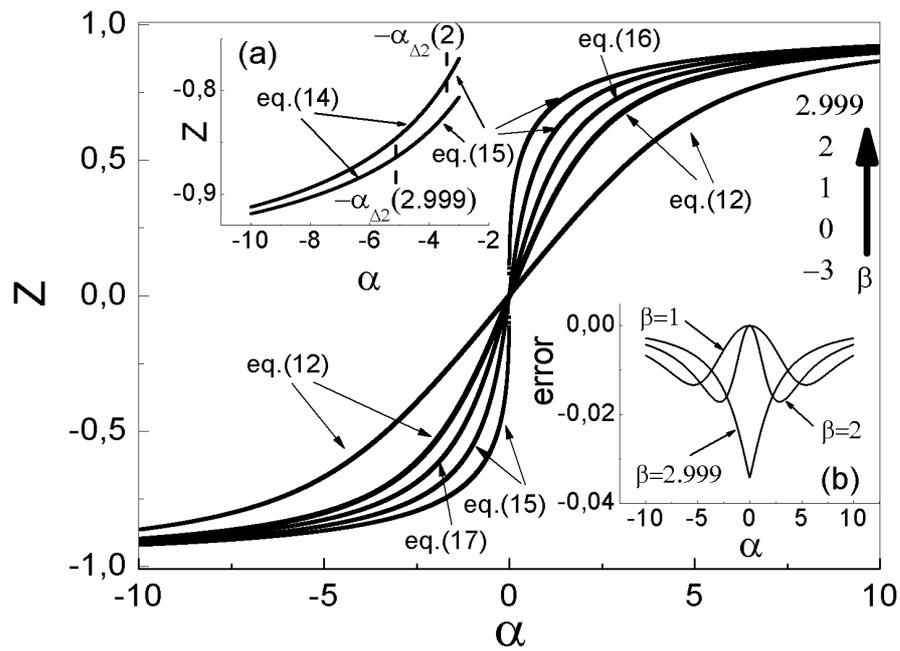

Fig. 3. The curves $z(\alpha)|_{\beta=\text{const}}$ for different values of $\beta$ satisfying the condition $\beta<3$. These values are indicated at the top right of the Figure. Inset (a): the portion of the curves for $\beta=2.999$ and $\beta=2$ on an enlarged scale. Inset (b): the relative error of solutions (12)-(17) with reference to the numerical solutions of (1) for the curves for $\beta=1$, 2 and 2.999. The numerical solutions of (1) were calculated by the iteration method [18] with the relative error of $10^{-8}$.

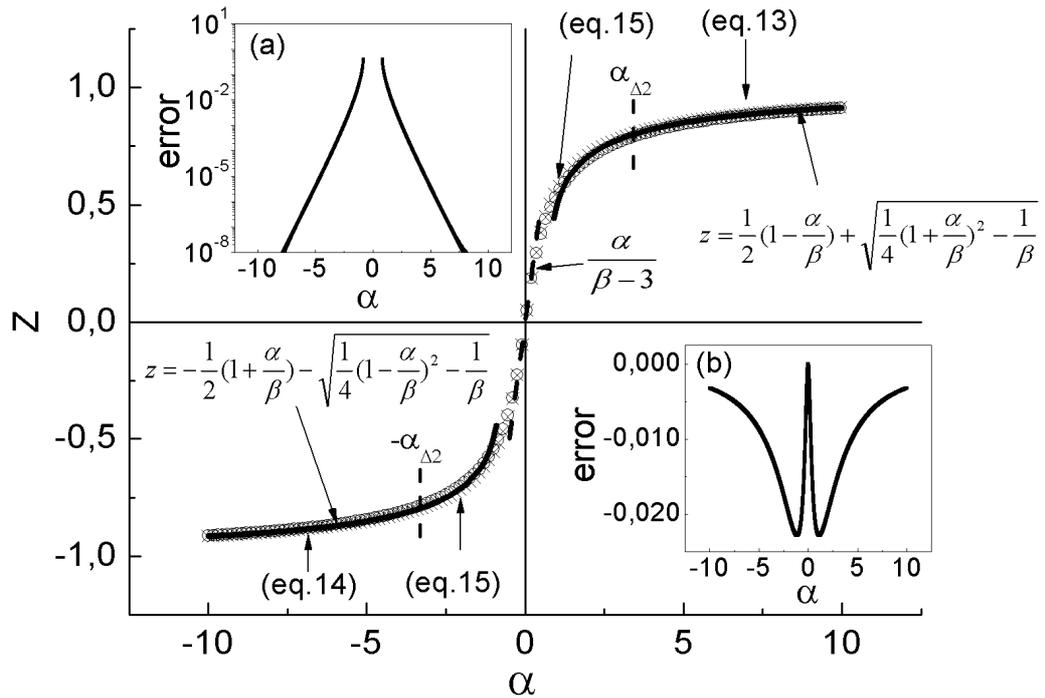

Fig. 4 The curves $z(\alpha)|_{\beta=2}$ calculated by (13)-(15) (crosses) and (8), (9) (solid lines). The circles denote the numerical solution of (1), which was calculated by the iterative method [18] with the relative error of $10^{-8}$. The dashed line corresponds to the tangent line at $\alpha=0$ according to equation (21). The inset (a) and (b) show the relative error of solutions (8), (9) and (13)-(15), respectively, with reference to the numerical solution of (1).

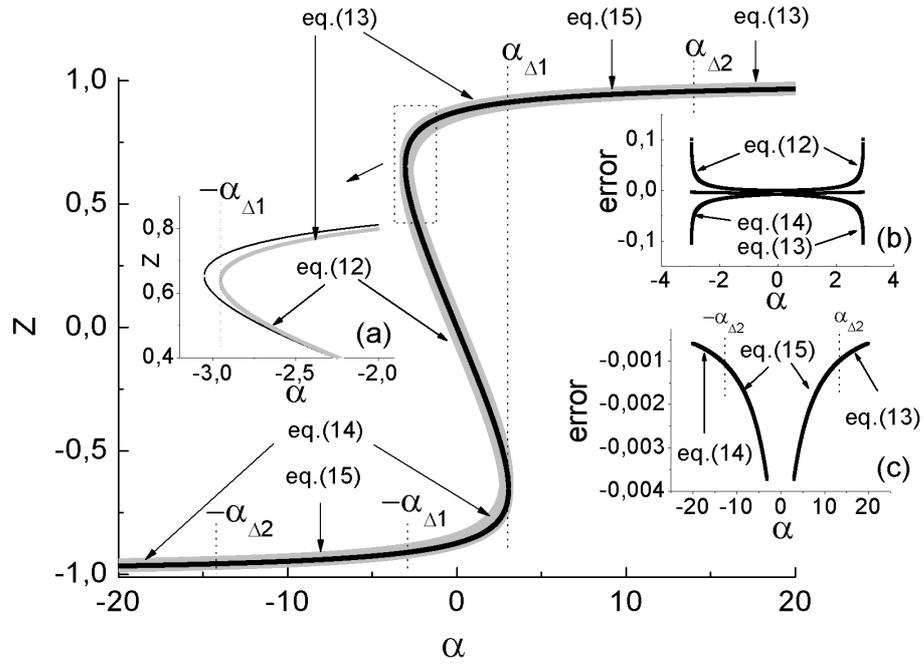

Fig. 5 The curve $z(\alpha)|_{\beta=9}$ (circles) calculated by (12)-(15). Inset (a): the portion of the curve near $\alpha = -\alpha_{\Delta 1}$, where $\left(\dfrac{\partial z}{\partial \alpha}\right)_\beta$ diverges. The solid lines represent the numerical solutions of (1) for $\beta=9$. Insets (b): the relative error with reference to the numerical solutions of (1) for (12)-(14) if $|\alpha|<\alpha_{\Delta 1}$. Inset (c): the relative error with reference to the numerical solution of (1) for (13) if $\alpha>\alpha_{\Delta 2}$, for (14) if $\alpha<-\alpha_{\Delta 2}$ and for (15) if $\alpha_{\Delta 2}>|\alpha|>\alpha_{\Delta 1}$. The numerical solutions of (1) were calculated by the bisection method [18] with the relative error of $10^{-8}$.

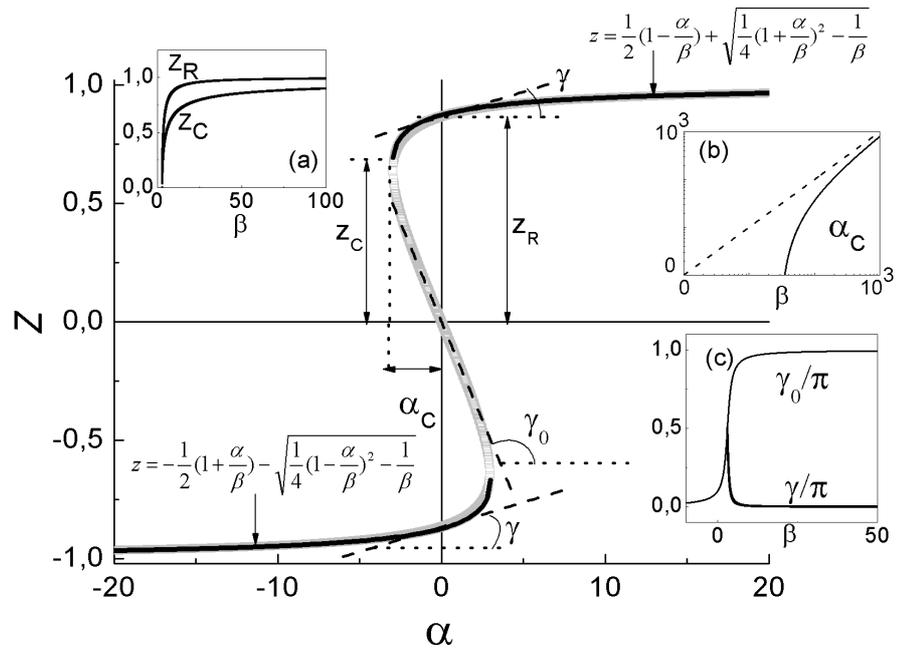

Fig. 6 The details of the "S"-shaped curve $z(\alpha)|_{\beta=9}$. The circles and the solid lines represent the solutions (12)-(14) and (8), (9), respectively. The symbol $z_C$ and $\alpha_C$ denote the modulus of the ordinate and abscissa, respectively, of the points at which $\left(\dfrac{\partial z}{\partial \alpha}\right)_\beta \to \infty$. The angles $\gamma_0$ and $\gamma$ define at $\alpha=0$ the slope of the curve (12) and (13), (14), respectively. Insets (a), (b) and (c) show the $\beta$-dependency of $z_C$ and $z_R$, $\alpha_C$, $\gamma$ and $\gamma_0$, respectively. The dashed line in inset (c) represents $\alpha=\beta$ as the asymptotic formula for $\alpha_C$ if $\beta\to\infty$.

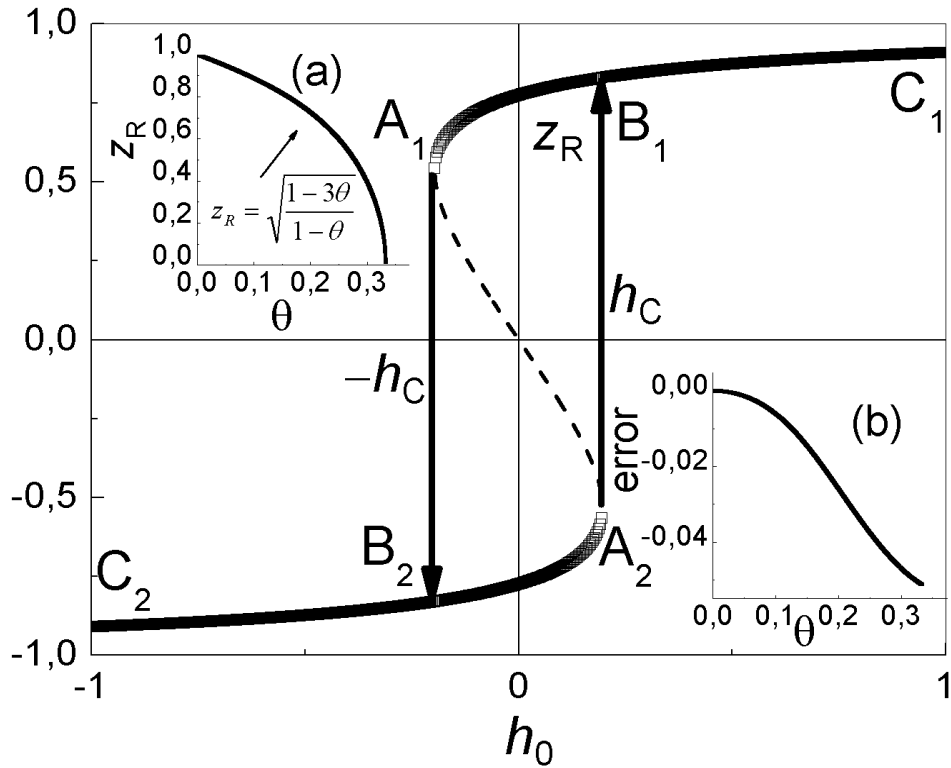

Fig. 7 The isothermal ($\theta=\theta_C/2$, $\beta=6$) curve of the normalized magnetization $z(h_0)$. The residual magnetization is denoted by $z_R$. The symbol $h_C$ represents the modulus of $h_0$ at which the transition $A_1 \to B_2$ or $A_2 \to B_1$ takes place. The curves $C_1B_1A_1$ and $C_2B_2A_2$ corresponds to continuous changing of the magnetization. Inset (a): the temperature dependence of $z_R$ (see (61)). The dashed line $A_1A_2$ represents unstable states. Inset (b): the relative error of $z_R$ with reference to the numerical solution of (1). The numerical solutions of (1) were calculated by the bisection method [18] with the relative error of $10^{-8}$.

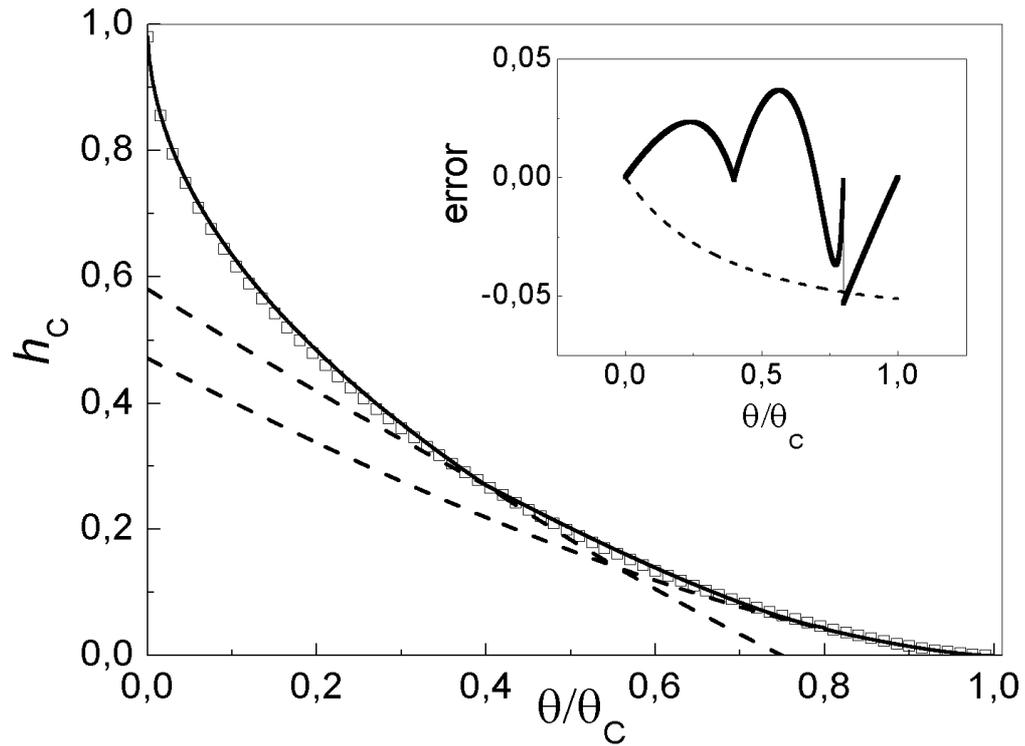

Fig.8 The temperature dependence (55) of the coercive force $h_C$ (solid line). The circles represent the values of $h_C$ calculated by means of the numerical solution of (1). The approximate dependences (56) and (57) are indicated by the dashed lines. The inset shows the relative error with reference to the numerical solution of (1) for relationship (55) (dashed line) and its approximations (56), (57) and (58) (solid lines) in the interval $\theta/\theta_C \in [\theta_F, 1]$, $[0, \theta_S]$ and $[\theta_S, \theta_F]$, respectively, where $\theta_S=2/5$, $\theta_F=4/5$.

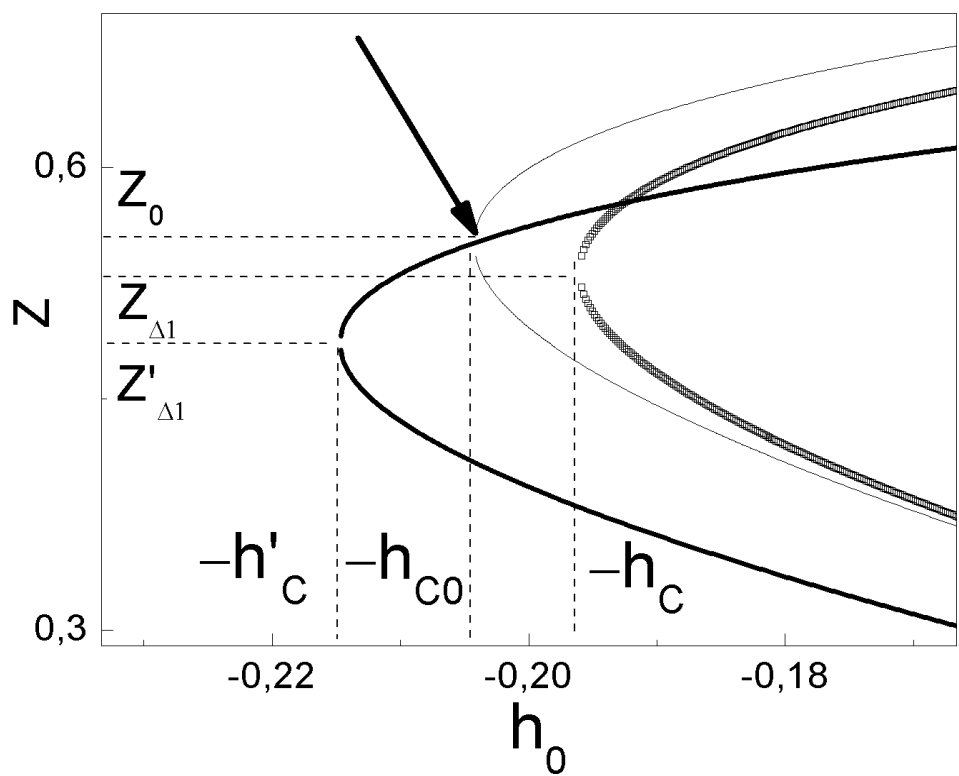

Fig. 9 The solutions of equations (5) (squares), (65) (thick lines) and (1) (thin lines) for $\beta=6$ ($\theta=\theta_C/2$). The arrow shows the point at which the derivative (64) diverges. This point determines the exact value $h_{C0}$ of the coercive force. The values of $h_C$, $h'_C$, and $z_0$ are defined by equations (54), (73), and (72), respectively. The symbols $z_{\Delta 1}$ and $z'_{\Delta 1}$ are introduced in analogy to $z_C$ in Fig.6. The value of $z_{\Delta 1}$ is determined by (26).

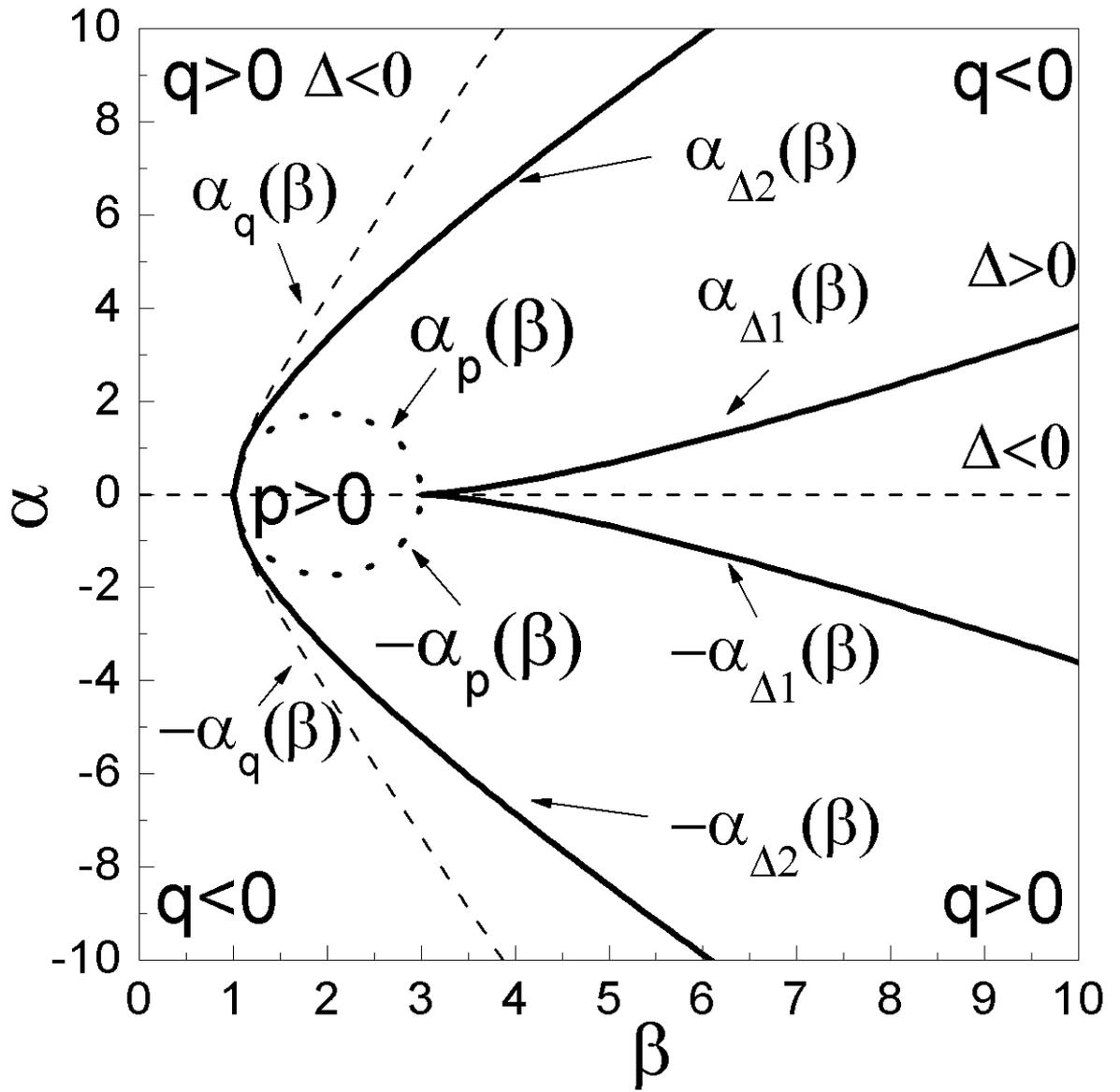

Fig. A The boundaries of the sign change in Δ (solid lines), *p* (dotted line) and *q* (dashed lines) (see (A15)-(A18)) in the (α, β) plane.